\documentclass[twocolumn,showpacs,preprintnumbers,amsmath,amssymb]{revtex4}

\usepackage{graphicx}
\usepackage{dcolumn}
\usepackage{bm}

\begin{document}

\newcommand{\ket}[1]{\ensuremath{|#1\rangle}}
\def\lbar{\ensuremath{\overline{l}}}
\newcommand{\ketbra}[1]{\ensuremath{|#1\rangle\langle #1 |}}
\def\be{\begin{eqnarray}}
\def\ee{\end{eqnarray}}

\title{The Second Quantized Quantum Turing Machine and Kolmogorov Complexity}

\author{Caroline Rogers}

\affiliation{Department of Computer Science, University of
Warwick CV4 7AL, UK}

\author{Vlatko Vedral}

\affiliation{School of Physics and Astronomy, University of Leeds
LS2 9JT, UK }

\affiliation{Institut fur Experimentalphysik, Universit\"{a}t Wien, Wien,
Boltzmanngasse 5 1090, Austria}

\begin{abstract}
The Kolmogorov complexity of a physical state is the minimal
physical resources required to reproduce that state. We define a
second quantized quantum Turing machine and use it to define
second quantized Kolmogorov complexity. There are two advantages
to our approach -- our measure of second quantized Kolmogorov
complexity is closer to physical reality and unlike other quantum
Kolmogorov
 complexities it is continuous. We give examples where second quantized and quantum Kolmogorov complexity differ.
\end{abstract}


\maketitle


Quantum physics, as far as we know, is the most accurate and
universal description of all physical phenomena in the universe.
If we therefore wish to speak about the complexity of some
processes or some physical states in nature, we need to use
quantum physics as our best available theory. An important
question is (in very simple terms), given a physical system in some state, how difficult
is it for us to reproduce it. If we wish to have a universal
measure of this difficulty (which applies to all systems and
states) a way to proceed is to follow the prescription of
Kolmogorov. First, we define a universal computer (which is
capable of simulating all other computers) and then we look for
the shortest input (another physical state) to this computer that
reproduces as the output the desired physical state. This way,
all the complexity is defined with respect to the same universal
computer, and we can thus speak about universal complexity. The
universal computer (such as a universal Turing machine) needs to
be fully quantum mechanical in order to capture all the
possibilities available in nature.\\
In this letter, we define a fully quantized Turing machine which
we use to define a fully quantum Kolmogorov complexity and apply
it to a number of different problems. Our approach is different
from others in that we consider indeterminate length input
programs whose expected length is our measure of complexity.
Others, whose work is discussed in detail below, have perhaps
avoided our approach, as allowing programs in superposition can
lead to a superposition of halting times. Since traditionally
computation is viewed as giving a deterministic output after a
fixed amount of time, having a superposition of halting times
seems to contradict the very concept of computation. We, on the
other hand, view the superposition of different length inputs
which can lead to a superposition of halting times as a necessity
dictated by a fully quantum mechanical description of nature.
Moreover allowing superpositions of different programs can lead to
a program which has on average a smaller Kolmogorov complexity
than when programs are restricted to having a variable but
determinate length (we give examples of this in this letter).
Since we want to know what is physically the shortest input which
produces a given output, we need allow the computation with
superpositions of different length programs.\\
This letter is organized as follows. First we discuss the concept
of a Turing machine and review previous work. Then we define the
fully quantum Turing machine (we call it second quantized for
reasons that will become apparent below). We use this second
quantized Turing machine (SQTM) to define the concept of second
quantized Kolmogorov complexity (SQKC). This notion is then tested
by applying it to various simple examples such as the average
length of two programs of different sizes and programs which halt
at a superposition of different times. We consider conditional
complexity and show that multiple copies of a quantum state can be
compressed asymptotically further using SQTM than on a standard
quantum Turing machine (QTM) given the number of copies $n$.


Quantum information is usually studied in $n$-particle systems
described by an $n$-fold tensor product Hilbert space $H^{\otimes
n}$ in which $n$ is fixed. In the second quantization $n$, the
number of particles, can exist in superposition. The corresponding
space which describes a system in the second quantization is the
Fock space
$H^*=\bigoplus_{n=0}^{\infty}H^{\otimes n}$.\\
A physical example of such a system is the quantized
electromagnetic field. The second quantized electromagnetic field
consists of a number of modes $\omega_i$ (labelled by frequency
and polarization) which can each be populated by a number of
photons $m_i+n_i$, $m_i$ in the horizontal polarization $H$ and
$n_i$ in the vertical polarization $V$.  The state of the system
can be written as: \vspace{-0.2cm}
\begin{equation}
\ket{\psi}=\ket{m_1}_{\omega_{1}, H}\ket{n_1}_{\omega_{1},
  V}\otimes\ket{m_2}_{\omega_{2}, H}\ket{n_2}_{\omega_{2},
  V}\otimes\ldots \vspace{-0.2cm}
  \end{equation}
  For the purposes of this letter, we will encode
qubits into the polarization degree of freedom, which then
restricts the number $m_i+n_i$ of quanta in the modes to either
$0$ (the vacuum state) or $1$. The modes are ordered in some
predefined way, and the initial contiguous sequence which are
occupied by a single photon before the first vacuum state
represents a quantum string. For example, the state \be
\ket{\psi}&=&\frac{1}{\sqrt{2}}\ket{1}_{\omega_1,H}\otimes\ket{0}_{\omega_1,V}\otimes\ket{0}_{\omega_2,H}\otimes\ket{1}_{\omega_2,V} \nonumber \\
&\otimes&\left((\ket{1}_{\omega_3,H}\otimes\ket{0}_{\omega_3,V})+
  (\ket{0}_{\omega_3,H}\otimes\ket{1}_{\omega_3,V})\right) \\
  &\otimes&\ket{0}_{\omega_{4},H}\otimes\ket{0}_{\omega_{4},V}\otimes\ldots
\ee can be written as a quantum string $
\ket{\psi}=\ket{01}(\ket{0}+\ket{1})/\sqrt{2}$, where a photon in
the horizontal or vertical polarizations represents a $\ket{0}$ or
$\ket{1}$ respectively. If a photon exists at a frequency in
superposition, the quantum string may be in a superposition of two
different lengths. For example,\vspace{-0.2cm} \be
\ket{\psi}&=&\frac{1}{\sqrt{2}}\ket{1}_{\omega_1,H}
\otimes\ket{0}_{\omega_1,V}\otimes
\ket{0}_{\omega_2,H} \nonumber \\
&\otimes& (\ket{0}_{\omega_2,V}+\ket{1}_{\omega_2,V})
\otimes\ket{0}_{\omega_3,H}\otimes\ket{0}_{\omega_3,V} \\
&\otimes&\ldots \vspace{-0.2cm} \ee which can be written as the quantum string $
\ket{\psi}=(\ket{01}+\ket{0})/\sqrt{2}$. Note that the
superposition of different program lengths immediately implies a
superposition of different times of computation. If a head moves
along the string from left to right, it will reach the end of the
string at different times directly proportional to the length.
Since variable length encoding naturally leads to the concept of
creation and annihilation of particles \cite{ral02}, we call the
Turing machine described in this letter the Second Quantized
Turing machine (SQTM).

\vspace{0.3cm} Now we briefly review similar work. Svozil
\cite{svo96} originally defined quantum Kolmogorov complexity on a
circuit based model and then Berthiaume, van Dam and Laplante
\cite{ber01} defined quantum Kolmogorov complexity \cite{vitanyi} on
a quantum Turing machine. \cite{svo96} and
\cite{ber01} both restricted inputs to be of variable but
determined length (i.e. quantum but not second quantized inputs).
Vitanyi \cite{vit01} also provided a definition of quantum
Kolmogorov complexity based on classical descriptions, where a penalty factor was added
depending on the accuracy of the classical description.
Mora and Briegel \cite{mor04,mor05}
defined the algorithmic complexity of a quantum state to be the classical Kolmogorov complexity
of describing a circuit that produces the quantum state (up to an error parameter $\epsilon$).

Gacs \cite{gacs} and Tadaki \cite{tad02} defined quantum
Kolmogorov complexity without reference to a computation device by
generalizing classical universal semi-measures to quantum
universal semi-POVM's. Tadaki \cite{tad04} went on to derive
$\Omega_Q$, a quantum generalization of Chaitin's halting
probability \cite{chaitin}. More recently, Benatti {\it et al} \cite{ben05} 
have given a quantum Brudno theorem and M\"{u}ller \cite{markuspaper} has 
given a detailed proof of the invariance theorem 
for Berthiaume {\it et al}'s complexity \cite{ber01}.

The standard models of quantum Turing machines were defined by
Deutsch \cite{deu85} and Berstein and Vazirani \cite{ber97}.
Nielsen \cite{nie97} also defined a programmable quantum gate
array. Nishimura and Ozawa \cite{har02} compared various models of
quantum computation. Various papers \cite{mye97, oza98, shi02,
miy02} have discussed how these models can halt coherently (all
computation paths halting simultaneously) when two programs which
halt at different times are used as input in superposition.
\cite{miy02} showed that it is undecidable whether a quantum
Turing machine halts coherently for a general input.

Schumacher and Westmoreland \cite{sch01}, Bostr\"{o}m and Felbinger
\cite{bos02} and Rallan and Vedral \cite{ral02} studied lossless
quantum compression with variable length quantum strings. One of the
authors \cite{losslessquantumcompression} studied exact lossless quantum compression and
universal lossless quantum compression with indeterminate length
quantum strings.

\vspace{0.3cm} Now we discuss the SQTM. The aim of developing an
SQTM is to fully quantize every aspect of the quantum Turing
machine (QTM) to produce a fully quantum computational model of
nature. Unlike a QTM, the SQTM's definition makes it explicit that it  can 
hold second quantized states on the tape(s) and so that it can halt at a superposition of different times.
\begin{description}
\vspace{-0.2cm} \item[Second Quantized Tape] The contents of the
tape including the input and output is allowed to exist in the
second quantization. States such as
$\ket{\psi}=\frac{1}{\sqrt{2}}(\ket{0}+\ket{10})$ are forbidden on
the QTM but allowed on the SQTM. \vspace{-0.2cm}
 \item[Second
Quantized Halting] Several papers \cite{mye97, oza98, shi02,
miy02} have pointed
  out issues in how quantum computations halt coherently (how all the
  computational paths halt simultaneously) on the quantum Turing
  machine \cite{ber97, deu85} and Nielsen's parallel gate array
  \cite{nie97}. The SQTM allows programs of different sizes to be
  input in superposition. These programs may halt at different times,
  however as natural processes may also halt at a superposition
  of different times, the SQTM allows this. Informally, the halting condition says that the SQTM comes 
  to a halt when all the computation paths of the working tape stop evolving.
\end{description}


The SQTM is made up of a two-way infinite working tape, an
environment tape, a transition function, a head which points to
the current cell being processed (perhaps in superposition) and a
current instruction being executed (perhaps also in
superposition). Each cell in the tape contains either a $\ket{0}$,
$\ket{1}$ or a vacuum state. The string held on the tape is made
up of the \ket{0}'s and \ket{1}'s before the first vacuum state
which can exist in a superposition of positions.

The head of the Turing machine contains two quantum states. The
first \ket{C}, contains the current superposition of cells being
processed, the second \ket{I}, the current instruction being
executed.  The transition of the Turing machine from one state to
the next is controlled by the transition function $\delta$ which
can act on the current and neighboring cells. At each step, the
head can move left, right or the current cell being scanned can be
modified and the internal state of the head can be changed with
the restriction that the whole operation of the Turing machine is
restricted to be a unitary operation. This unitary operation is
computable and can be computably written down classically. Using
this classical description, there exists a universal SQTM capable
of simulating any other SQTM with a constant length description.

The environment tape is defined in the same way as the working
tape, except that it receives no input and gives no output. The SQTM can continue to carry out
computations on the environment tape (for example, counting up to
infinite) to ensure that its operation is unitary. The proposed halting scheme for
the SQTM (though others may be possible) is as follows. On input
$\ket{\psi_{I}}$, the SQTM is said to halt in the state
$\ket{\psi_{F}}$ if the contents of the working tape converges to
\ket{\psi_{F}} computably, i.e. if there exists a computable
sequence of integers $t_1$, $t_2$, $\ldots$ such that for all
$t\geq t_i$ steps, the SQTM has executed $t$ steps and the working
tape of the SQTM (with the state of the other parts of the SQTM
traced out) is in state $T_{SQ}^{t}(\ket{\psi_{I}})$ where $
 |\langle T_{SQ}^{t}(\ket{\psi_{I}}) \ket{\psi_F} |\leq 1/2^i
$. The restriction of the sequence $t_i$ to be computable allows
the SQTM to be simulated by some deterministic classical Turing
machine to compute a classical description of $\ket{\psi_F}$ with
arbitrary accuracy, ensuring that the SQTM halts in a computable
state when given computable input. The halting condition also
guarantees that there exists some SQTM which on input \ket{i} will
halt in the state \ket{\psi_{F}} with a fidelity of $1/2^i$.
\vspace{0.3cm}


The classical Kolmogorov complexity of a string is defined as the
shortest classical description of that string with respect to a
universal Turing machine. Kolmogorov complexity gives a universal
measure of the complexity of a string and has many applications
such as the incompressibility method, data compression, universal
induction and absolute information.  Following from classical
Kolmogorov complexity, there are several possible ways to define
second quantized Kolmogorov complexity (SQKC) \cite{kolmogorov,vitanyi}.
\begin{description}
\vspace{-0.2cm}
\item[The combinatorial approach] The
combinatorial approach is to
  study the number of dimensions used to described a quantum state
  $\ket{\psi}$. The combinatorial approach is similar to the
  restriction of input to be variable but determined length, which is used by Svozil
  \cite{svo96} and then Berthiaume {\it et al} \cite{ber01} in their definitions
of quantum Kolmogorov complexity. \vspace{-0.2cm}
\item[Algorithmic probability] This approach is to
  define Kolmogorov complexity in terms of probability theory avoiding
  the need to refer to a Turing machine. Tadaki \cite{tad02, tad04} and Gacs
  \cite{gacs} took this approach for quantum Kolmogorov complexity. \vspace{-0.2cm}
  \item[Algorithmic Approach] The algorithmic approach is to consider
    the complexity of a string as the length of the shortest algorithm
    or description which describes that string, the idea being that
    simpler strings can be described by shorter descriptions. For
    example, $x=1111111111$ is very simple and regular and has a very
    short description as ``ten ones'' whereas $y=1000111010$ is
    apparently much more random   and it seems difficult to find a short
    description of $y$.  A second quantized description may be in a
    superposition of many different lengths. It is the average length of the description
    which is used to measure the second quantized Kolmogorov
    complexity of a string giving the average physical resources used
    to describe the state.
\vspace{-0.2cm}
\end{description}
We believe the average length is the appropriate measure of
complexity because it in some sense corresponds to the average
energy of the input \cite{ral02} and energy is frequently the most
crucial physical resource. Unlike the combinatorial approach, it
leads to a continuous measure of complexity which is a much more
natural physical quantity. We now study the
algorithmic approach and by SQKC we mean the average length
complexity used in the algorithmic approach.

Let $\ket{\psi}=\sum_{i=0}^{\infty}\alpha_{i}\ket{i}$ be a second
quantized state. The average length \cite{bos02} $\lbar$ of
$\ket{\psi}$ is \vspace{-0.2cm}
\be \lbar(\ket{\psi})&=&\sum_{i=0}^{\infty}
|\alpha_i|^2 l(i) \vspace{-0.2cm}
\ee 
The SQKC of a state \ket{\psi} is \vspace{-0.2cm}
\be
C_{SQ}(\ket{\psi})&=&\inf_{U_{SQ}(\ket{\phi})=\ket{\psi}}
\lbar(\ket{\phi}) \vspace{-0.2cm}
\ee 
where $U_{SQ}$ is a universal SQTM. The SQKC
$C_{SQ}$ is invariant up to an additive constant factor (this
invariance theorem can be proved in the same way as the classical
invariance theorem\cite{vitanyi}).

A simple example of SQKC, which is different from QKC, is to try
to find the shortest string which can describe the state $
\ket{\psi_n}=(\sqrt{2}\ket{0}+\ket{1^{\otimes n}})/\sqrt{3} $ for
$n>0$. A shortest program to describe \ket{\psi_n} is
$\ket{p_{\psi_n}}=(\sqrt{2}\ket{0}+\ket{n})/\sqrt{3} $ for $n>0$.
\ket{p_{\psi_n}} can be input into an SQTM $T_{SQ}$ defined by $
T_{SQ}\ket{n}=\ket{1^{\otimes n}}$ and $T_{SQ}\ket{0}=\ket{0}$ to
produce \ket{\psi_n} as output. The SQKC of \ket{\psi_n} is then
at most the average length of \ket{p_{\psi_n}} plus a constant
factor $O(1)$ which comes from describing $T_{SQ}$ on the
universal machine $U_{SQ}$. The SQKC of \ket{\psi_n} is
at most \vspace{-0.2cm} \be
C_{SQ}(\ket{\psi_n})&\leq&\lbar(\ket{\phi_{\psi_n}})+O(1) \\
&=&\frac{2}{3}O(1)+\frac{1}{3}(\log(n)+O(1)) \\
&=&\frac{1}{3}\log(n)+O(1) \vspace{-0.2cm}
\ee 
On the other hand, the quantum Kolmogorov complexity of $1^n$ is $\log(n)+O(1)$ which is the number of qubits used in a fixed
length description of \ket{\psi_n}. Thus the SQKC of \ket{\psi_n} is multiplicatively smaller than the quantum Kolmogorov complexity
of \ket{\psi_n}. Notice also that if the amplitude $1/\sqrt{3}$ is changed to $\alpha$ and 
$|\alpha| \to 0$ then $C_{SQ}(\psi_n)\to C_{SQ}(0)$, which is not true of the quantum Kolmogorov
complexity defined by Berthiame {\it et al} \cite{ber01}.

The \ket{\psi_n} defined above can also be used to show how halting at a superposition of different times is used in
 SQKC.
As above, let $ \ket{\psi_n}=(\sqrt{2}\ket{0}+\ket{1^{\otimes
n}})/\sqrt{3}$. Suppose that an SQTM can compute \ket{0}
in $\tau$ steps. For sufficiently large $n$, \ket{1^{\otimes n} } takes more than $\tau$ steps to compute. Thus, for large $n$,
the SQTM must be allowed to halt at a superposition
of different times to describe \ket{\psi_n}. Thus it is by allowing the SQTM to halt at a the number of
steps in superposition that we can describe states which are in a
superposition of different lengths.

The conditional SQKC of a string \ket{\psi} given \ket{\phi}
is the complexity of \ket{\psi} given $\ket{\phi}$ as input to the universal
SQTM.
\vspace{-0.2cm}
\begin{equation}
C_{SQ}(\ket{\psi}|\ket{\phi})=\inf_{U(\ket{\phi},p)}\lbar(p) \vspace{-0.2cm}
\end{equation}
The SQKC of $n$ copies of a state \ket{\psi} assuming that $n$ is
known in advance is also a non-trivial example of how SQKC differs
from QKC. Taking $\ket{\psi}=\alpha\ket{0}+\beta\ket{1}$,
$\ket{\psi}^{\otimes n}$ can be expanded out into the symmetric basis
as
\vspace{-0.2cm}
\be
\ket{\psi}^{\otimes n}&=&(\alpha\ket{0}+\beta\ket{1})^{\otimes n} \\
&=&\sum_{i=0}^n \alpha^i \beta^{n-i} S(i,n) \vspace{-0.3cm}
\ee
where $S(i,n)$ is the
sum of strings containing $i$ $0$'s and $n-i$ $1$'s.

Berthiaume {\it et al} \cite{ber01} showed that $\ket{\psi}^{\otimes
  n}$ can be described using $\log(n)$ qubits using a fixed length code
to encode each $S(i,n)$. On the other hand, using a variable length code to encode
$S(i,n)$ as a string of $\log{n \choose i}$ qubits, $\ket{\psi^{\otimes
    n}}$ can be encoded using $S(\rho_{\psi})$ qubits asymptotically
where $S$ is the von Neumann entropy and $\rho_\psi=\sum_{i} |\alpha^i
\beta^{n-i}|^2 |S(i,n)\rangle\langle S(i,n)|$ is the state of
\ket{\psi} after a measurement has been carried out in the
$\{S(i,n)\}_{i}$ basis.
Thus as $n$ grows,
the number of bits used to describe $\ket{\psi^{\otimes n}}$ with an
SQTM can be far fewer than are used by a QTM.

We have defined a second quantized Turing machine which formally
models information processing in second quantized physical systems
and addresses how computational processes halt (the halting of the
SQTM). We have used this SQTM to define second quantized
Kolmogorov complexity which improves on the previous measure, quantum
Kolmogorov complexity, in that it is continuous and therefore a
much more natural physical quantity.\\
 Other measures of
information such as von Neumann entropy and its variants are
widely used for studying various properties of quantum systems
such as entanglement, distances between systems, etc. These other
information measures can be seen as computable approximations of
(second quantized) Kolmogorov complexity. Kolmogorov complexity is
much more powerful. For instance, if there is some underlying
mechanical description of a set of experiments (e.g. the Bohm
interpretation) then the measurement outcomes
of the experiment may have a lower Kolmogorov complexity than
if the measurement outcomes are completely random while the
entropy of the outcomes are the same
in cases.\\
 There are many other possible applications of the
SQTM. One example is Maxwell's Demon.
An attempt was made (and subsequently resolved) to show that the second law can be violated in the classical setting \cite{set97, cav90, zur99}, 
using the fact that a sequence of particles might have a very compact description while their joint entropy is high. Maxwell's demon can be considered in its fully quantum setting
with the help of the SQTM. As we have argued SQKC can be smaller
than QKC, which may help the demon in trying to violate the second
law in the quantum case. This question is left for future research.\\
In this paper, we have studied a few examples of SQKC. Classical
Kolmogorov complexity is a well-studied area with a plethora of
applications in many areas of physics. We hope that the future
development of SQKC will lead to a plethora of tools for studying
quantum systems and that the differences in classical and second
quantized Kolmogorov complexity will yield further insights into
the differences between classical and quantum systems.
\vspace{-0.1cm}

The authors thank Charles Bennett, Johann Summhammer,  Karl Svozil and Andreas Winter for useful discussion.
The authors thank the EPSRC
for financial support. V.V. also
thanks the European Union and the British Council in Austria.

\vspace{-0.4cm}

\bibliography{kolmogorov4}

\end{document}